\documentclass[preprint]{aa}
\usepackage{natbib}
\usepackage{graphicx} 
\usepackage{hyperref}
\usepackage{txfonts}
\def\it{\sl}
\def\degs{\ifmmode ^{\circ}\else$^{\circ}$\fi}
\def\amin{\ifmmode ^{\prime}\else$^{\prime}$\fi}
\def\asec{\ifmmode ^{\prime\prime}\else$^{\prime\prime}$\fi}

\def\degs{\ifmmode ^{\circ}\else$^{\circ}$\fi}
\def\amin{\ifmmode ^{\prime}\else$^{\prime}$\fi}

\def\eqalign#1{\null\,\vcenter{\openup1\jot \m@th
   \ialign{\strut\hfil$\displaystyle{##}$&$\displaystyle{{}##}$\hfil
   \crcr#1\crcr}}\,}
\usepackage{color}

\begin{document}

\title{Dynamical Analysis of the 3:1 Resonance in the $\upsilon$ Andromedae System}

\offprints{Carlos E. Chavez \\  \email{carlosepech@astrosen.unam.mx}}

\author{Carlos E. Chavez
\and  Mauricio Reyes-Ruiz
\and  Hector Aceves
}

\institute{
Instituto de Astronom\'{\i}a, Universidad Nacional Aut\'{o}noma
de M\'{e}xico, Apdo. Postal 106, Ensenada, Baja California, 22800 M\'{e}xico.
}

\abstract{In this paper we study the dynamics of the $\upsilon$ Andromedae planetary system proposed by Curiel et~al.~(2011). 
We focus on the study of the 3:1 Mean Motion Resonance (hereafter MMR) between $\upsilon$ Andromedae--d and the 
recently discovered $\upsilon$ Andromedae--e (hereafter $\upsilon$ And--d and $\upsilon$ And--e).}{Numerical 
simulations of the dynamics of the four planet system are conducted.}{The previously 
reported apsidal resonance between  $\upsilon$ And--c and  $\upsilon$ And--d is confirmed. In addition, we find 
that $\upsilon$ And--d and  $\upsilon$ And--e are also in an apsidal resonance condition.}{Our results further indicate
that the $\upsilon$ Andromedae planetary system configuration is in the middle of a stability island in 
the semimajor axis-eccentricity domain. Additionally, we performed numerical integrations of the
planetary configuration over 500 Myr and found it to be stable.}{We conclude that, within the uncertainties in the value 
of the orbital parameters, it is likely that $\upsilon$ Andromedae planetary system will remain stable for a long timescale.}

\keywords{planets and satellites: dynamical evolution and stability -stars: individual: $\upsilon$ And}

\maketitle

\section{Introduction}

About $\sim 10\%$ of the exoplanets known to date are in multiple planet systems (http://exoplanet.eu). 
This is likely to be a lower limit since there is still a strong observational bias towards the detection 
of planets with short periods. Therefore, some of the extrasolar planets detected so far may have planetary 
companions not yet detected on distant orbits (e.g. \citet{Correia2009}).

Of the exoplanets discovered in multiple planet systems, several cases have been found to be in mean 
motion resonances (MMR). This is important because resonances tend to stabilize the orbits of the planets 
involved, \citet{Murray} section 8.3. Of the cases discovered to date, the 2:1 mean motion resonance 
is the most common (e.g. HD 73526, HD 82943, HD 128311, GJ 876, Kepler 9 and HD 37124), 
but there are other configurations such as the the 3:2 MMR in HD 45364, 
the 3:1 MMR in HD 75732 and the 5:1 in HD 202206. In our Solar System, the 3:2 MMR between Neptune and Pluto (albeit
not a {\it bona-fide} planet) that allows the latter to be in a high eccentricity and high inclination on an orbit that
crosses the orbit of Neptune, such that these two objects will never collide (e.g. \citet{Varadi}). In the context 
of the so-called {\sl Nice} model, it is believed that in the early Solar System, Jupiter and Saturn crossed the 2:1 
resonance, leading a significant rearrangement of the general architecture of our planetary system 
(\citet{Tsiganis}, \citet{Morbidelli} and \citet{Gomes}).

It has been pointed out recently by \citet{Correia2009}, that the study of the dynamics of mean motion resonances 
of two or more planets interacting in a system, offers the opportunity to constrain and understand the process of 
planetary formation and evolution, since these resonances most probably arise from planetary migration.

The $\upsilon$ Andromedae (hereafter $\upsilon$ And) planetary system was the first multiple, extrasolar planetary 
system discovered orbiting a solar-type star \citep{Butler}. The system is known to harbour three extrasolar planets
with masses ranging from 0.69 to 14.57 M$_{\rm J}$, with M$_{\rm J}$ being the mass of Jupiter. Recently, \citet{Curiel} have found a fourth planet orbiting 
$\upsilon$ And named, as is the convention, $\upsilon$ And--e. It is the pourpose of the work presented in this paper
to analyse the dynamics of this new planet to determine whether it is in MMR with $\upsilon$ And--d and, hence, in a likely stable 
configuration.

The paper is organized as follows. In Section 2 we review the general characteristics of the central star $\upsilon$ And 
and the known orbiting planetary system. In Section 3, we perform a global frequency analysis of the two planets possibly 
involved in the 3:1 resonance, namely, $\upsilon$ And--d and $\upsilon$ And--e. Additionally, we present results on the long term 
evolution of the system. Section 4 focuses on the apsidal resonance between $\upsilon$ And--c and $\upsilon$ And--d and 
Section 5 focuses on the apsidal resonance between $\upsilon$ And--d and $\upsilon$ And--e. Finally, 
we summarize our results and offer our conclusions in section 5.


\section{$\upsilon$ Andromedae system characteristics}
\label{upsandcharact}
$\upsilon$ And is a bright F8V star with a mass of 1.3 $M_{\odot}$ and stellar radius of 1.56 $R_{\odot}$ 
\citep{Butler}. The distance to the star is estimated to be about 13.47 pc, that is 43.93 lyrs 
\citep{Perryman}. The estimated age of the star is 5 Gyrs \citep{Baliunas}, and its 
rotational period is between 9 and 12 days (\citet{Baliunas}, \citet{Ford}).
$\upsilon$ And was the first multiple exoplanetary system detected around a main sequence star.
It was reported as being a triple planetary system by \citet{Butler}. The estimated masses and orbital 
paramenters for each of the planets around $\upsilon$ And are listed in Table \ref{table1}. Several studies have been made about 
the origin and dynamical stability of the triple planetary system (\citet{Ford2005}, \citet{Chiang2001}, 
\citet{Rivera} and references therein). The main conclusion of these studies, considering 
only the first three planets discovered are: \citet{Ford2005} found out that planet--planet scattering with a fourth lost planet could explain the high eccentricities of $\upsilon$ And--c and $\upsilon$ And--d. \citet{Chiang2001} found out that the apsidal resonance between $\upsilon$ And--c and $\upsilon$ And--d is observed for mutual inclinations, between these two planets,  smaller than $20^{\circ}$. \citet{Rivera} concluded that there are stability regions for test particles around the 3:1 and 5:1 MMR, and that for $a>7.5$ AU all test particles were stable for at least $10^7$ yrs. They reported additionally that test particles just outside the 1:3 MMR with $\upsilon$ And--d experience large oscillations reaching eccentricities up to the range of 0.2 to 0.3; but these particles are protected from close approaches with $\upsilon$ And--d by the e--$\omega$ mechanism.
\begin{table}
\caption{Properties of the planetary companions in $\upsilon$ And (from \citet{Curiel}).}      
\label{table1}      
\centering                                      
\begin{tabular}{c c c c}          
\hline\hline                        
Planet & M (M$_J$) & $a$ (AU) & $e$ \\    
\hline                                   
    b & 0.6876(44) & 0.05922166(20) & 0.02150(70) \\      
    c & 1.981(19) & 0.827774(15)   & 0.2596(79) \\
    d & 4.132(29) & 2.51329(75)        & 0.2987(72) \\
    e & 1.059(28) & 5.24558(67)      & 0.00536(440) \\
\hline                                             
\end{tabular}
\end{table}
\begin{table}
\caption{Fundamental frequencies calculated from the nominal solution}      
\label{table2}      
\centering                                      
\begin{tabular}{c c c }          
\hline\hline                        
 & Frequency ($^\circ$yr) & Period (yr) \\    
\hline                                   
    $n_{d}$ & 0.285194 & 1262.3 \\      
    $n_{e}$ & 0.0947104 & 3801.06   \\
    $g_{d}$ & 0.0000257259 & 38871.3      \\
    $g_{e}$ & 0.00109351 & 914.484   \\
    $l_{\theta_{3}}$ & 1.52241 & 236.467   \\
    $l_{e_{4}}$ & 0.415535 & 866.353   \\
\hline                                             
\end{tabular}
\end{table}
Recently, \citet{Curiel} have discovered a fourth planet orbiting the system on the basis of
a refined fit for the radial velocity data. Its properties are also described
in Table \ref{table1}. \citet{Curiel} propose that the system is close to a 3:1 MMR resonance. 
The initial estimate for the period of the fourth planet ($\upsilon$ And--e) is 3848.86 days, 
and the period of the third planet ($\upsilon$ And--d) is 1276.46. The ratio between the two 
periods is 3.02, very close to being in exact resonance. In a previous study, \citet{Rivera} find that there is an island of stability in the semimajor axis-eccentricity
parameter domain, just outside the external values corresponding to the 3:1 MMR. 
\citet{Rivera} use a large collection of test particles to sample 
the possible location of new planets.

\begin{figure}[t]
\includegraphics[width=8.5cm,height=12.5cm]{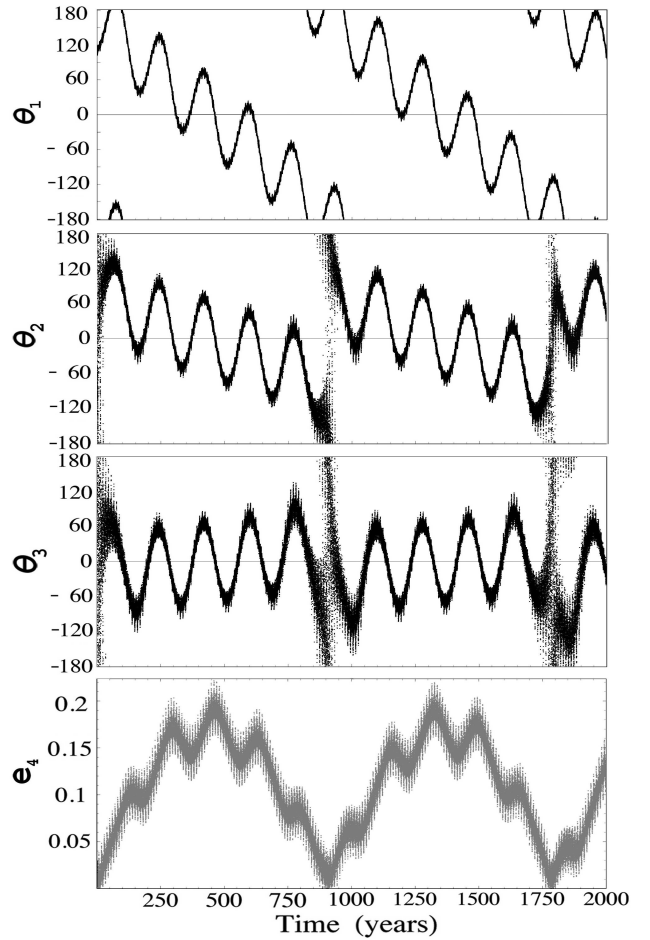}
\caption{Time evolution of the three resonant angles for the 3:1 resonance between $\upsilon$ And--d 
and $\upsilon$ And--e.  The angles $\theta_{1}$ and  $\theta_{2}$ circulate, 
while $\theta_{3}$ librates. The bottom panel shows the oscillation in the
time evolution of the eccentricity of $\upsilon$ And--e (called here $e_{4}$)}
\label{fig1}
\end{figure}
\begin{figure*}[t]
\includegraphics[width=17.0cm]{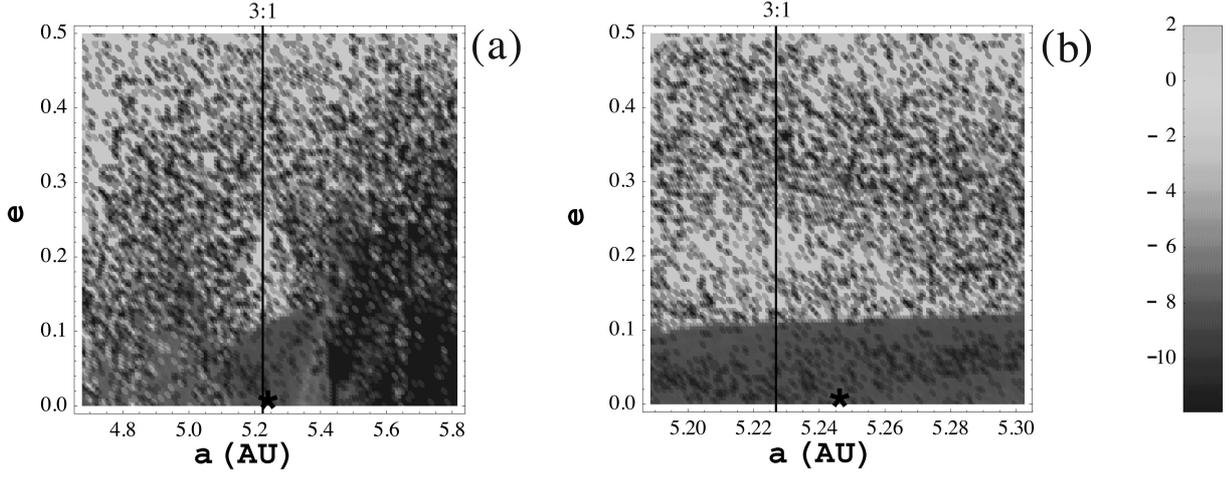}
\caption{Stability analysis of the $\upsilon$ And system as a function of semi-major axis, $a$, 
and eccentricity for planet-e, $e$. The logarithm of the stability parameter $D$ is shown on a gray scale map.
The lighter colors correspond to the more unstable systems. The log--value $+2$ for the parameter $D$ 
corresponds to the ejection of planet-e. The $\bigstar$ symbol corresponds to the nominal solution reported
by \citet{Curiel}. Panel b) is a close-up, in $a$, of the wider domain shown in panel a) of the 
parameter space around the nominal solution.} 
\label{fig2}
\end{figure*}

\begin{figure*}[t]
\includegraphics[width=17.0cm]{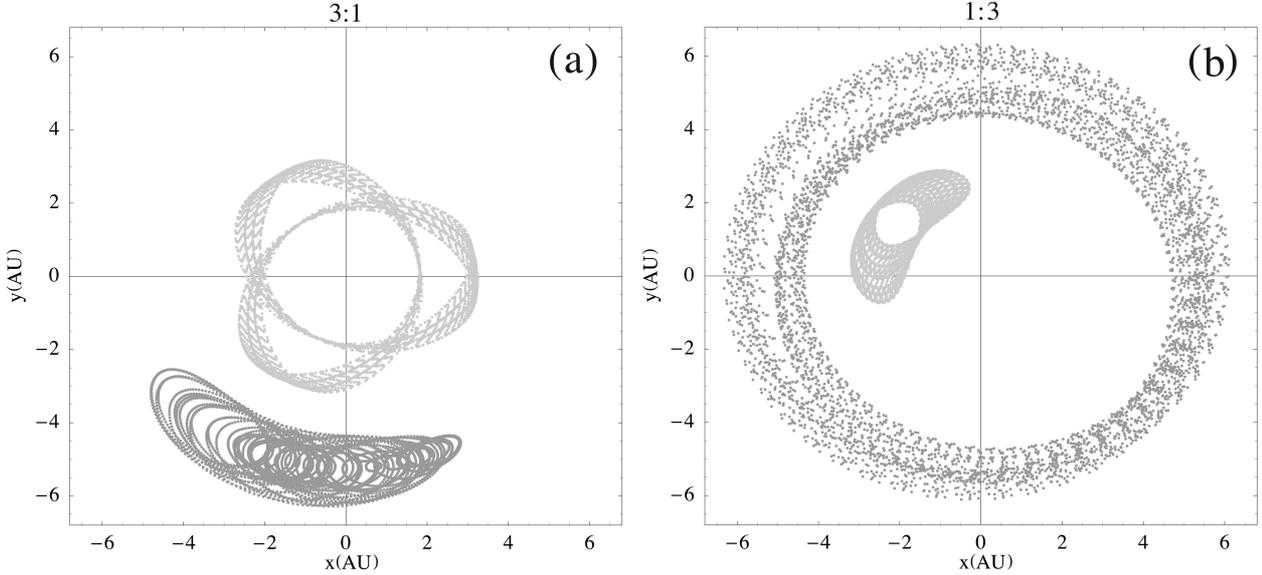}
\caption{Orbits in rotating reference frames for 500 yrs for the $\upsilon$ And system. Panel (a) shows 
the reference frame with the angular velocity $M_{c}$ in which the 3:1 MMR resonance is observed. 
Panel (b) shows the same points but in a reference frame that is rotating with $M_{d}$, 
where we observe the 1:3 MMR resonance}
\label{fig3}
\end{figure*}

\begin{figure}[t]
\includegraphics[width=\columnwidth,height=12.5cm]{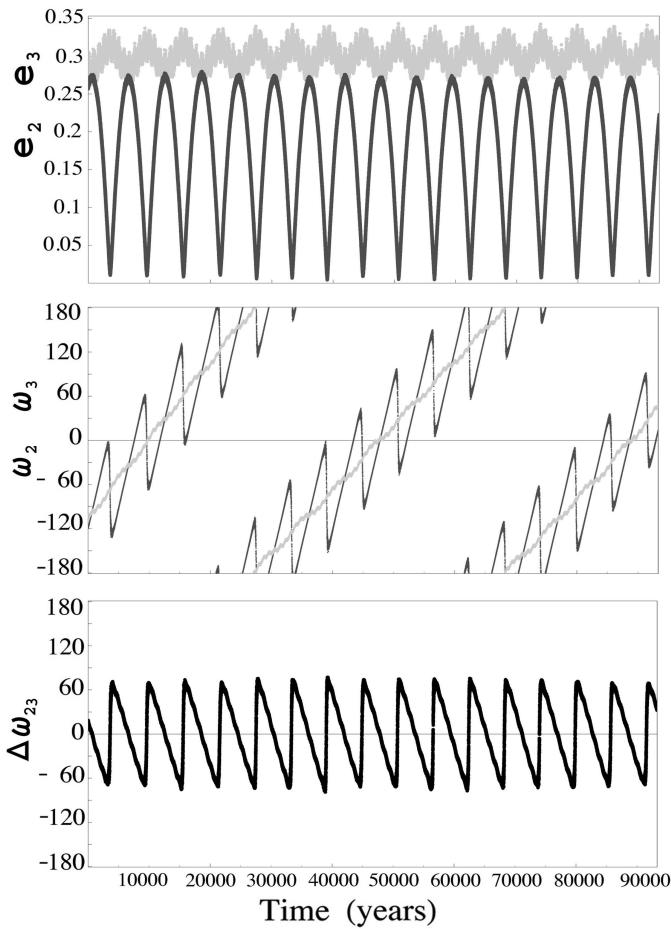}
\caption{Apsidal resonance between planet-c and planet-d. The two eccentricities are anti-correlated as shown in the top panel. The eccentricity of planet c (here $e_{3}$) gets close to zero periodically. We show in the middle panel the argument of the pericentre of the two planets, it can be noticed that they tend to follow each other.
Finally, in the bottom panel we show the difference between the arguments of pericentre ($\Delta \omega_{23}$) of planets c and d. As we can notice this angle is librating around zero.   
}
\label{fig5}
\end{figure}
\begin{figure}[t]
\includegraphics[width=\columnwidth,height=12.5cm]{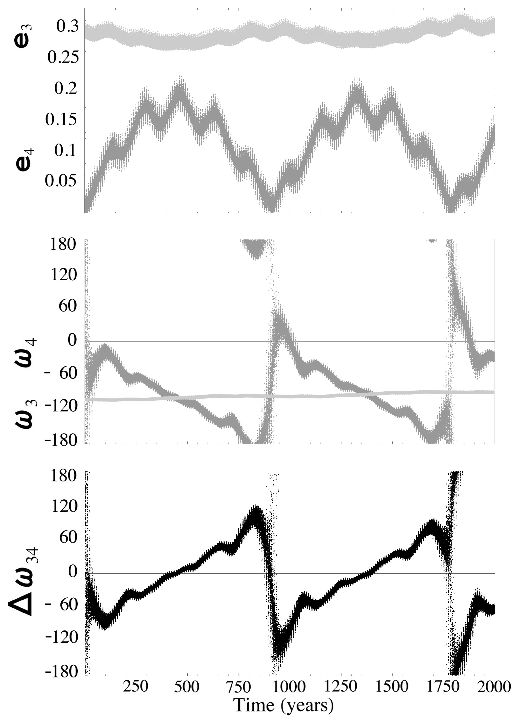}
\caption{Apsidal resonance between planet-d and planet-e. The two eccentricities of planets d and e (shown here in light grey and dark grey, respectively) are anti-correlated. The middle panel shows the arguments of pericentre of these two planets, it is possible to notice that they follow each other. Finally, in the bottom panel is shown the difference between these two arguments of pericentre ($\Delta \omega_{34}$), it is possible to notice that this angle is librating around zero.}
\label{fig6}
\end{figure}

\section{Orbital stability}
\label{orbstability}

In this section we analyse the dynamical stability of a planetary system characterized by the orbital parameters 
reported in Table \ref{table1}. We refer to this set of parameters for the $\upsilon$ And planetary system as the nominal solution. There are different procedures to check the stability of a system using the frequency analysis developed by \citet{Laskar1993} (e.g. \citet{Marzari2005}). We follow the approach taken by \citet{Correia2005}, \citet{Correia2009} and  \citet{Couetdic2010} to analyse the stability of the 
planetary configuration.

\subsection{The 3:1 resonance}

To test whether the $\upsilon$ And system is trapped or not in the 3:1 MMR resonance,
we perform a frequency analysis of the nominal solution computed over 10 Kyrs. 
Since the orbit of the inner planets of the system is well constrained, 
we fix the orbital parameters of $\upsilon$ And-b, $\upsilon$ And-c and $\upsilon$ And-d
according to Table \ref{table1}. We also assume the orbits of all planets are coplanar.
A series of initial conditions for the eccentricity ($e$) and semi-major ($a$) axis of
planet e are constructed to test the stability of the region around 
the nominal solution. 
  
For each initial condition, the orbits of the planets are integrated 
over 10 kyrs using the hybrid integrator included in the
$Mercury$ 6 code \citep{Chambers}. For most of the integration, $Mercury$ uses a 
mixed-variable symplectic integrator \citep{Wisdom1991} with a time step 
approximately equal to a fiftieth ($\approx$1/50) of the Keplerian orbital 
period of the closest planet (in this case $\upsilon$ And--b). During close 
encounters, $Mercury$ uses a Bulirsch-Stoer integrator with an accuracy 
parameter of $10^{-12}$.
The stability of the orbit corresponding to each set of ($a, e$) values
is measured using the frequency analysis introduced by \citet{Laskar1990} and \citet{Laskar1993}. 
According to \citet{Laskar1993} the difference ($D$) in the value of the fundamental 
frequency of the motion of the planet under consideration, obtained over 
two consecutive time intervals 
is a measure of the secular stability of the trajectory. 

In addtion, we also identify strongly unstable systems as those in which: 
1) two planets collide, 2) a planet hits 
the star (if its astrocentric distance is $> 0.005$ AU), or 3) a planet is 
ejected from the system (assumed to occur if the planet travels beyond an
astrocentric distance of 100 AU).

The three possible resonant arguments of the 3:1 resonance according to 
\citet{Murray} p.491, with $\Omega = 0.0$ because we consider 
coplanar orbits, are the following:
\begin{equation}
\theta_{1}=3 \lambda_{e} - \lambda_{d} - 2 \omega_{d} ,
\label{theta3}
\end{equation}
\begin{equation}
\theta_{2}=3 \lambda_{e} - \lambda_{d} - \omega_{e} - \omega_{d}
\label{theta3}
\end{equation}
\noindent and
\begin{equation}
\theta_{3}=3 \lambda_{e} - \lambda_{d} - 2 \omega_{e} ,
\label{theta3}
\end{equation}
where $\lambda_{d}$, $\omega_{d}$ and $\lambda_{e}$, $\omega_{e}$ are the mean 
longitude and argument of the pericenter of planet e and d, respectively.
Fig. \ref{fig1} shows the time evolution of the three possible resonant angles
and the time evolution of the eccentricity of planet-e, in order to make a 
direct comparison of its periodicity with the resonant angles. We
find that $\theta_{1}$ and $\theta_{2}$ are circulating while $\theta_{3}$ 
is librating around zero. This behaviour means that the two planets are closer to the 
resonance defined by $\theta_{3}$.  

It is clear form Figure \ref{fig1} 
 that $\theta_{2}$ and $\theta_{3}$ have periodic and sudden changes with 
periods of 866.353 yrs. (we found this period using a Fourier analysis), this 
period coincides with the period of approximately 866.353 years that can be seen in the 
oscillation of the eccentricity of planet-e about zero. This behaviour can be explained 
by the fact that $\omega_{e}$ is not well defined for circular orbits.
Therefore $\theta_{3}$ is not circulating every 866.353 yrs., but rather librating 
and every period that planet e comes back to zero this resonant angle suffers 
the sudden changes observed in Fig. \ref{fig1}.

The fundamental frequencies of the system are the two mean motions 
(known also as mean angular velocities) $n_{d}= {{d \lambda_{d}} \over {d t}}$ 
and $n_{e}= {{d \lambda_{e}} \over {d t}}$, the two secular frequencies of 
the pericenter $g_{d}= {{d \omega_{d}} \over {d t}}$ and 
$g_{e}= {{d \omega_{e}} \over {d t}}$, and the libration frequency of the 
resonant argument $l_{\theta}= {{d \theta_{3}} \over {d t}}$. The values of each one of these are shown in Table \ref{table2}.

\subsection{Stability analysis}

To analyse the stability of our nominal solution, we perform a global 
frequency analysis \citep{Laskar1993} in its vicinty 
(Fig.  \ref{fig2}) (a) and (b), similarly as it has been done by \citet{Correia2005}, \citet{Correia2009} and \citet{Couetdic2010}.
For the two planets that are possibly involved in the the resonance (d \& e), 
the system is integrated using a two-dimensional mesh 
of 11,165 initial conditions. We vary the semi-major axis and eccentricity 
of planet e, and keep the orbital parameters of the other planets at their 
nominal values. 

We integrate each case numerically over 10 kyrs, and then 
we compute the stability indicator $D$. This stability indicator is computed 
by comparing the variation in the measured mean motion over two 
consecutive $T=5$ kyr intervals. The parameter $D$ is defined as:

\begin{equation}
D= {{{| n_{e}-n'_{e} |}  \over T}} ,
\label{parameterD}
\end{equation}

\noindent where $n_{e}$ and $n'_{e}$ are the mean motion of planet e in the 
first 5 kyr and last 5 kyr of the integration, respectively; the units of $D$ 
are deg/yr$^{2}$. This parameter $D$ is a measurement of the chaotic diffusion of the trajectory. 
It should be close to zero for a regular solution and it has a high value for strong chaotic motion \citep{Laskar1993}.
We also performed 10 integrations of 1 Myrs and found that in the $\upsilon$ And system, a
value of $D < 10^{-7}$ is required for regular motion (comparing the $D$ of stable 
and unstable integrations). 

The left panel of Figure \ref{fig2} ($\mathbf{a}$) shows the results of our mesh of 11,165 numerical 
simulations in the domain of semimajor axis, $a \in [4.7,5.8]$ AU, and eccentricity $e \in [0,0.5]$. 
From Fig. \ref{fig2} ($\mathbf{a}$) we see that there is an island of stability 
around the nominal solution, shown here with the symbol $\bigstar$ (located at $a=5.24558$ AU).
The region does not seem to contain ejected particles (designated with $D= +2$ in this plot). 
We show also the location of the 3:1 MMR as a vertical line at $a=5.22672$ AU. 
At this scale the location of the nominal solution and the location of the 3:1 MMR are very 
close to each other. Therefore we decided to zoom-in around the region of the nominal solution, and explore it numerically. 

We performed another set of 11,615 integrations around the stability island, 
focusing on the parameter domain $a \in [5.18,5.30]$ AU and $e \in [0,0.5]$. 
Our results are shown in the right panel of Fig. \ref{fig2} ($\mathbf{b}$).
The shape of the stability island around the 3:1 resonance is easily appreciated
on this scale. We notice that most of the systems explored in the $a$--$e$ mesh 
are unstable above $e=0.1$. The nominal solution reported in \citet{Curiel} 
has $a=5.24558$ AU and $e=0.00536$ (shown again with $\bigstar$). That puts the system 
right in the middle of the stability island. This result support the idea that 
the solution found by \citet{Curiel} is stable and robust, as suggested by these 
authors. Also, it appears that the only stable zone that exists in the vicinity of the 
nominal solution is the zone protected by 3:1 resonance.

In Fig. \ref{fig3} ($\mathbf{a}$ and $\mathbf{b}$) we show the evolution of 
$\upsilon$ And--d and $\upsilon$ And--e over 500 yr in the rotating reference 
frame of the inner and outer planet, respectively. Since the system is so 
close to the 3:1 resonance the relative position of the two planets is repeated, 
and their minimum distance in our simulation was $2.47$ AU. The paths of the two 
planets in this rotating reference frame shows the relationship between the 
resonance and the frequency of conjunctions (see \citet{Murray}, p. 325). 
In this particular reference frame, every three orbits of $\upsilon$ And--d corresponds to one 
orbit of $\upsilon$ And e. Figure \ref{fig3} also shows the libration of each 
planet around its equilibrium position. Since there are two other planets on 
the system $\upsilon$ And, the equilibrium position precesses and does not maintain its path, in the rotating reference 
frame, for a much longer time.

\subsection{Orbital evolution}

The previous stability analysis supports the idea that the $\upsilon$ And system is trapped 
in the island--like stability zone associated with the 3:1 resonance, and that it is very likely 
to be stable on long term evolution. To test this suggestion we carried out a long 
term numerical integration of the nominal solution. Using the $Mercury$ 6 code with the 
same parameters as described above, we integrated the orbits of the four planets 
for 500 Myrs. We found that, as suggested by the study of \citet{Curiel}, 
these orbits are stable in the long term.

Due to the intense resonant gravitational interaction the three planets involved (c, d \& e) exhibit significant changes in their eccentricities, 
as can be seen in Figs. \ref{fig5} and \ref{fig6}. 
\section{Apsidal resonance between $\upsilon$ And-c and $\upsilon$ And-d}

In this section we report on the apsidal resonance that was observed in previous 
research on $\upsilon$ And (e.g. \citet{Curiel}, \citet{Chiang2001}, \citet{Rivera2000}, 
\citet{Lissauer2001}, \citet{Chiang2002}, \citet{Barnes2004}, 
\citet{Michtchenko2004}, \citet{Nagasawa2005}, etc). 

The secular apsidal resonance is shown in Fig. \ref{fig5}. We find that the 
eccentricities $e_{d}$ and $e_{4}$ are anti-correlated as well, that is, 
when $e_{d}$ is maximum $e_{4}$ is minimum and vice versa. We show $\omega_{c}$ 
(dark gray) and $\omega_{d}$ (light gray) in the middle panel 
of Fig. \ref{fig5}. The difference between these two angles is shown in the 
bottom panel of Fig. \ref{fig5}, it shows that $\Delta \omega_{cd}$ oscillates around 
zero with an initial amplitude of $\sim 77^{\circ}$ and a short period 
of $6 \times 10^{3}$ yrs. Therefore showing that planets $c$ and $d$ are
indeed in apsidal resonance.

The physics of the apsidal resonance is the following. The eccentricity 
of $\upsilon$ And--c and and $\upsilon$ And--d are quite large 
($e_{c}=0.2596$ and $e_{d}=0.2987$, Table \ref{table1} of 
\citep{Curiel}) and their semi-major axis are not too different 
from each other ($a_{c}=0.827774$ AU and $a_{d}=2.51329$ AU). In principle, 
they could be as close to each other as $D_{\rm min} = a_{d} (1-e_{d})-a_{c} (1+e_{c})=0.719906$ AU, 
but this never happens. As it is possible to notice in Fig. \ref{fig5},  
when the two orbits are aligned (that is $\Delta \omega_{cd}=0.0$) 
the eccentricity of planet-d is maximum, but at the same time the eccentricity 
of planet-c is reduced to values $e_{c} \sim 0$. This helps to avoid 
close encounters between the two planets, hence helping to stabilize the system,
then the two planets never get closer than 0.9 AU.

\section{Apsidal resonance between $\upsilon$ And-d and $\upsilon$ And-e}

In addition to planet--c and planet--d being in apsidal resonance, 
we find that planet--d and planet--e are in apsidal resonance as well. 
This was suggested by \citet{Curiel}, and here we give the details 
about this resonance for the first time. Figure  \ref{fig6} shows this 
apsidal resonance in action. We find that $e_{d}$ and $e_{4}$ are anti-correlated.
We also find that the maximum eccentricity that planet--e can achieve 
is quite big ($e_{max} = 0.233105$). Therefore, in principle, it is possible 
that planet--e and planet--d can get close to each other 
$D_{\rm min} =a_{4} (1-e_{4})-a_{d} (1+e_{d})=2.051726$ AU, but owing 
to the apsidal resonance, they never get closer than $D_{\rm min}=2.47421$ AU.

\section{Summary and conclusions}

We have carried out a stability analysis for the 4--planet system around $\upsilon$ And 
by means of a frequency analysis \citet{Laskar1993} and long term numerical simulations.
On the basis of our results, we find evidence that $\upsilon$ And--d and $\upsilon$ And-e 
are in  a 3:1 mean motion resonance, as suggested by previous studies. The nominal 
solution found by \citet{Curiel} is located in the middle of a island of stability
in the $a$--$e$ paramenter space. 

The $\upsilon$ And system is found to be a rich dynamical system, with three of the 
planets discovered to date interacting strongly via resonances. As described in the paper,
$\upsilon$ And e interacts via the 3:1 MMR with $\upsilon$ And d, and additionally they 
are in apsidal resonance. This prevents them from having close encounters. 
On another hand, the apsidal resonance reported by previous authors between $\upsilon$ And--c and 
$\upsilon$ And--d  is still present when $\upsilon$ And--e is considered. This means 
that the three planets are interacting via apsidal and MMR resonances. 

The nominal solution of \citet{Curiel} is here proved to be stable for 500 Myrs, 
this along with our results of the stability analysis done via the global frequency analysis,
allows us to conclude that the nominal solution is robust and stable.

Finally the reported eccentricity of $\upsilon$ And--e is at 
the moment very close to zero ($e=0.00536$). However, according to our results, it should 
increase to $e=0.016$ in approximately 10 years. If it is possible to measure this variation 
after such period, it will provide an important confirmation of the dynamical properties 
described in this work. Our results further predict that the eccentricity of $\upsilon$ And--e
will reach its maximum value of $0.2$ in around 450 years.

\acknowledgements
CC thanks the CONACyT postdoctoral program for its financial support. MRR acknowledges 
support from PAPIIT-UNAM project No. 109409. CC acknowledges 
support from CONACyT grant 128563. HA acknowledges 
support from PAPIIT-UNAM project IN109710.

\bibliography{fs_3b}

\end{document}